\documentclass[%
 reprint,
superscriptaddress,
groupedaddress,
preprintnumbers,
nobibnotes,
 amsmath,amssymb,
prb,
citeautoscript,
]{revtex4-1}

\usepackage{graphicx}
\usepackage{dcolumn}
\usepackage{bm}

\begin{document}


\title{Manifestation of Fermi edge singularity in the cotunneling regime}

\author{Artem Borin and Eugene Sukhorukov}
\affiliation{
 D\'epartement de Physique Th\'eorique, Universit\'e de Gen\`eve - CH-1211 Gen\`eve 4, Switzerland}

\date{\today}

\begin{abstract}
The Fermi edge singularity (FES) is a prominent manifestation of the Coulomb interaction. It can be observed in a controllable way by studying the transport through a quantum dot (QD), which is electrostatically coupled to the leads. In this paper we study how FES affects higher-order tunneling processes (cotunneling). To address this problem we use the bosonic description of the electrons in the leads, which naturally accounts for the Coulomb interaction. We report the multiscale dependence of the current through the QD on the energy of the QD level and on the bias between the leads obtained for the limit of large and small bias. The new universal powers are determined by the scattering phases due to the interaction of electrons in the leads with the charge on the QD.
\end{abstract}

\maketitle

\section{Introduction}

In contrast to the systems with the Fermi liquid like behavior, the electron-electron interactions become large in QD based devices. The Coulomb blockade is the most prominent manifestation of this  effect.\cite{CB} Its main feature is the appearance of the gap in the QD excitation spectra due to the finite energy required in order to add electron or hole to a QD. This energy is of the order of charging energy $E_C=e^2/2C$, where $e$ is the electron charge, and $C$ is the capacitance of the dot, and it can be controlled by the gate voltage. If in addition the interaction of the charge on the QD with metallic leads is taken into account, this results in the power-law dependence of the tunneling density of states at the energy close to the Fermi level. This effect is analogous to the one that leads to the singularity in X-ray absorption spectra in metals\cite{Xray} and is often referred to as the Fermi edge singularity. It has been extensively studied both experimentally\cite{experiment} and theoretically
\cite{theory1,theory2,theory3,theory4,theory5} in various systems.

Because of the large charging energy the dominant contribution to the transport through a QD is sequential tunneling,\cite{CB} i.e., the electron (hole) enters the dot only if the previous one has left it. This leads to the correlation of incoming and outgoing currents, which can be seen, e.g., in the suppression of the zero-frequency noise power.\cite{noise} The Coulomb blockade is not very  sensitive to the size of the QD, which only affects the value of the charging energy. Therefore, any mesoscopic QD demonstrates Coulomb blockade effect at sufficiently small temperatures. 

In contrast, the FES is more delicate phenomenon. It arises only if the number of transport channels of the QD is of the order of one.\cite{Xray} To be more precise, the exponent of the power-low energy dependence of the density of states at low energies is inversely proportional to the number of scattering channels. Since the large QDs are typically coupled to a large number of scattering channels, the FES effect is suppressed in such systems. It has been recently proposed to circumvent this difficulty by attaching large QDs to the quantum Hall channels,\cite{Anya} as it has been recently implemented experimentally.\cite{Heiblum}

Recent breakthrough on the theoretical side, namely, the development of the non-equilibrium bosonization approach,\cite{Eugene} enabled one to study the FES phenomenon far away from equilibrium.\cite{Iurii} One of the key results of this study is the universal dependence of the tunneling current on the both parameters: the energy of the level on the dot and the power of the non-equilibrium noise, which is controlled by an additional voltage source and provides the second energy scale. Here we propose an alternative approach, where the second energy scale is introduced by keeping the leads at equilibrium. Namely, we focus on the cotunneling regime away from the QB resonance,  where the sequential tunneling is suppressed, and the transport is dominated by simultaneous tunneling of two electrons or holes  via the dot.\cite{cotunneling} Such a process depends on two energy scales: the energy of the level on the QD and the bias between contacts.

Using the same approach that was developed in Ref.\ [\onlinecite{Iurii}] we obtain the universal power-law behavior of the tunneling current in two limits: in the case of small bias between the contacts, and in the case where the Fermi level in one contact is close to the energy level on the QD. We obtain new power-law exponents that depend on  the scattering phases of the electrons in the contacts, thus establishing the  connection to the ``classical'' FES effect. 

The rest of the paper is organized as follows: In Sec.\ II we introduce the model of a QD tunnel coupled to one-dimensional electron channels and present the Hamiltonian of the system using the bosonization technique. In Sec.\ III we present the formal perturbation theory to second order in tunneling, derive the expression for the cotunneling current, and discuss its analytical structure. In Sec.\ IV we obtain new FES exponents and discuss their physical significance. Finally, in Sec.\ V we present conclusions.

\section{Model}
In order to grasp the main features of the effect, we model the transport contacts to the QD by one-dimensional electronic channels. This model provides an effective description of metallic leads,\cite{Matveev} and can also be applied to quantum Hall systems,\cite{Wen} such as in the recent experiment [\onlinecite{Heiblum}]. It is well known,\cite{bosonization} that one-dimensional electronic systems can be described either in terms of the electrons or in terms of collective excitations (plasmons). FES is a non-perturbative effect in the electron-electron interactions. Since the Hamiltonian of one-dimensional channels, when expressed in plasmon fields, preserves its quadratic form in the presence of the interactions, we adopt this approach to study the FES. 

\begin{figure}[h]
	\centering
		\includegraphics[scale=0.3]{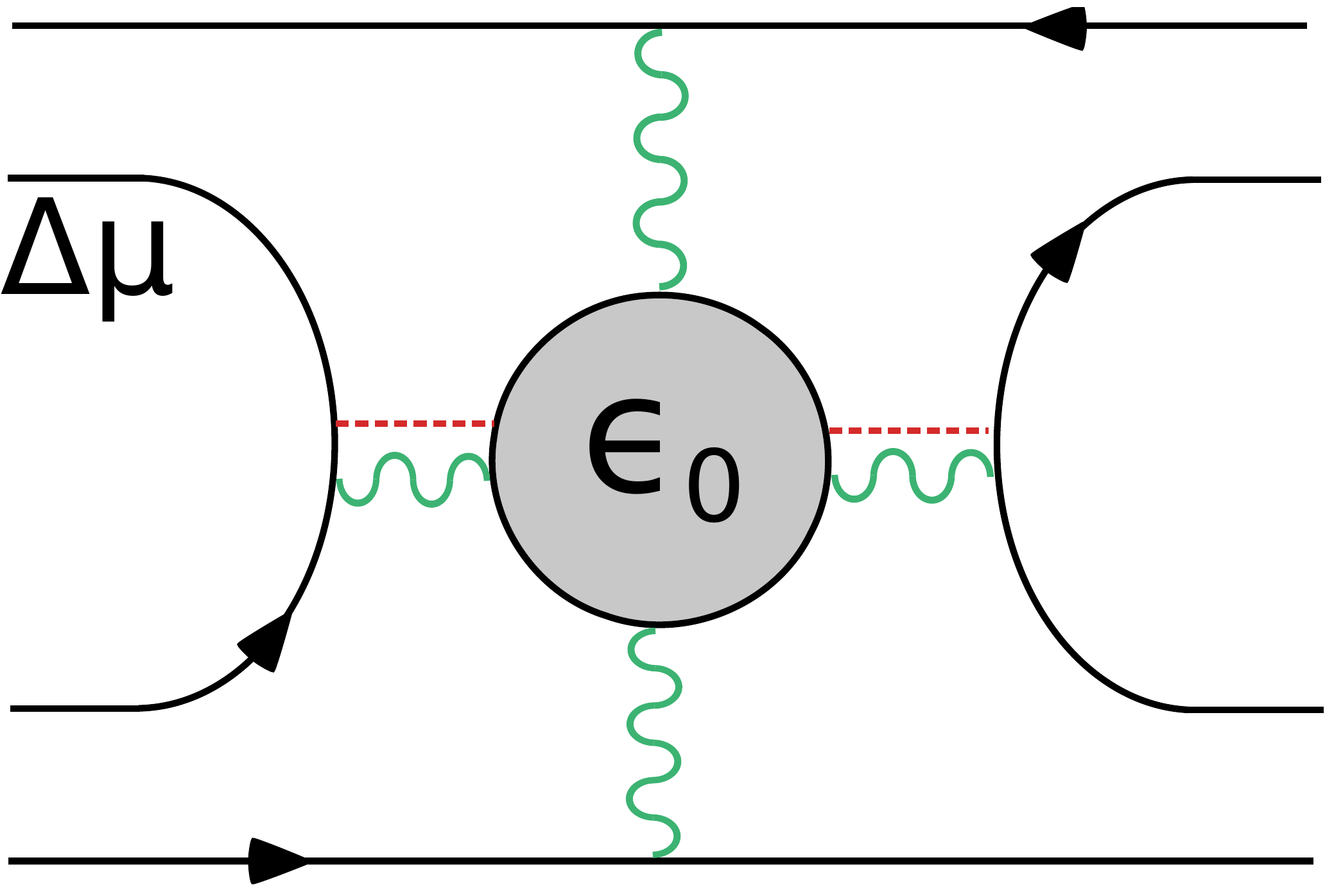} 
	\caption{A particular example of a more general system considered in this paper is schematically shown. This set up has been experimentally studied in Ref.\ [\onlinecite{Heiblum}] using QH edge states at filling factor $\nu=2$. The QD is coupled by the Coulomb interaction (green line) to the surrounding transport channels. The electrons tunnel from the left voltage-biased channel to the right grounded channel through the virtual states of the QD.}
\label{fig:setup}
\end{figure}

In this paper we concentrate on the regime of the elastic cotunneling,\cite{cotunneling} which can be realized in relatively small dots with few levels and relatively low biases. Thus, the minimal model for the FES  includes tunneling coupling of the dot to two electron channels, and Coulomb coupling to arbitrary number of channels (see Fig. \ref{fig:setup}). The Coulomb interaction should be considered non-perturbatively, while tunneling, in contrast, is the smallest perturbation. 
The corresponding Hamiltonian can be written as follows:
\begin{equation}
\mathcal{H}=H_0+H_d+H_i+H_t,
\end{equation}  
where the Hamiltonian of the electronic channels reads
\begin{eqnarray}
\nonumber H_0=\pi\int dx\sum_{a}v_{ a}\rho_a^2(x)\\ +\frac{1}{2}\int dxdy\sum_{ab}\rho_a(x)V_{ab}(x,y)\rho_b(y).
\end{eqnarray}
Here, the first part describes the free propagation of the charge densities $\rho_a(x)$ with the speeds $v_{ a}$, and $a$ enumerates channels.  The second part accounts for the density-density interactions characterized by the Coulomb potentials $V_{ab}(x,y)$. For the particular set up shown in Fig.\ \ref{fig:setup}, experimentally studied in Ref.\ [\onlinecite{Heiblum}], only left and right channel, denoted by $L$ and $R$ are coupled to the QD by tunneling.

The QD Hamiltonian can be written as $H_d=\epsilon_0 d^\dagger d$, where $d$ is the electron annihilation operator, and $\epsilon_0$ is the bare single-particle energy of the QD. 
The effect of FES arises due to the Coulomb interaction between the electron on the QD and those in the channels. The corresponding term in the Hamiltonian is given by
\begin{equation}
\label{inter}
H_i=d^\dagger d \int dx\sum_a U_a(x)\rho_a(x),
\end{equation}  
where $U_a$ are the Coulomb potentials. We note, that the potentials $U_a$ and $V_{ab}$ do not need to be specified because of the universality of FES, as we demonstrate below. 

Finally, the tunneling Hamiltonian transfers  electrons between the channels in the vicinity of the QD and the QD level, $H_t=d^\dagger\sum_a\tau_a\psi_a(0)+h.c.$, where $\psi_a(x)$ are the electron annihilation operators in the channels, and  $\tau_a$ are the amplitudes of tunneling. To make the connection between the electron and plasmon descriptions, we follow the standard bosonization procedure \cite{bosonization} and introduce the set of bosonic fields $\phi_a(x)$, which are defined by $\rho_a(x)=\frac{1}{2\pi}\partial_x \phi_a(x)$. They satisfy standard commutation relations $[\partial_x\phi_a(x),\phi_b(y)]=2\pi i\delta_{ab}\delta(x-y)$. The electron operators are related to the bosonic fields by $\psi_a(x)\propto e^{i\phi_a(x)}$. Therefore, the tunneling Hamiltonian can be written as 
\begin{equation}
H_t=d^\dagger\sum_a\tau_a e^{i\phi_a(0)}+h.c.\, .
\end{equation}

As a first step, we follow the procedure outlined in Ref.\ [\onlinecite{Iurii}] and apply the unitary transformation $U\equiv e^{iS}$ that removes the interaction term (\ref{inter}) in the total Hamiltonian. Here
\begin{equation}
S=d^\dagger d \int dx \sum_a \sigma_a(x)\phi_a(x),
\end{equation}
and the functions $\sigma_a(x)$ are determined by the integral equation 
\begin{equation}
U_a(x)+\int dx'\sum_b V_{ab}(x,x')\sigma_b(x')=0.
\label{screening}
\end{equation}
On one hand, such a choice of the functions $\sigma_a(x)$ follows from the requirement of the cancellation of the term $H_i$ in the Hamiltonian. On the other hand, the functions $\sigma_a(x)$ turn out to be equal to electron densities accumulated in the channels to screen the extra unite charge on the dot. From this point of view, the Eq.\ (\ref{screening}) simply states, that all the channels are grounded (see, however, the discussion below).

 After the introduced above unitary transformation the tunneling Hamiltonian acquires the following form:
\begin{equation}
\label{tun}
\tilde{H}_t=d^\dagger\sum_a \tau_a e^{i\int dx \sum_b \sigma_b(x)\phi_b(x)} e^{i\phi_a(0)}+h.c. \, .
\end{equation}
It is natural to assume, that the screening charges are accumulated in the vicinity of the point $x=0$. Thus, the simplification arises in the low-energy limit, where the wave length of plasmons exceeds the size of the QD, and leads to the universal behavior attributed to the FES. In this limit, we can approximate the fields in Eq.\ (\ref{tun}) $\phi_a(x)\approx\phi_a(0)$, which in our particular case leads to the expression
\begin{equation}
\tilde{H}_t=A_L+A_R+h.c.,
\end{equation}
with the elementary tunneling operators
\begin{subequations}
\label{A}
\begin{align}
A_L&=\tau_L d^\dagger e^{i\phi_L-i\sum_{a}\eta_a\phi_a},\\
A_R&=\tau_R d^\dagger e^{i\phi_R-i\sum_{a}\eta_a\phi_a},
\end{align}
\end{subequations}
and $\eta_a\equiv-\int dx \sigma_a(x)$ being the absolute values of the charges accumulated in the channels as a result of screening of the extra electron on the QD.

Finally, it is worth mentioning, that the unitary transformation shifts the energy level at the dot Hamiltonian: $H_d\to \tilde{H}_d=\epsilon d^\dagger d$, where
\begin{equation}
\label{remorm}
\epsilon=\epsilon_0+\sum_a\int dx U_a(x)\sigma_a(x).
\end{equation}
This shift, obviously, arises due to the Coulomb interaction of an electron on the dot with the induced charge densities in the channels.

\section{Perturbation theory}

We start with the standard Fermi golden rule expression for the cotunneling current at zero temperature:
\begin{equation}
\label{gamma}
I=2\pi \sum_{m}|\langle m|\hat{T}|0\rangle|^2\delta(E_m-E_0),
\end{equation}
where $|m\rangle$ are the eigenstates of the total Hamiltonian in the absence of tunneling. Since the first-order sequential tunneling processes are forbidden due to the energy gap in the cotunneling regime, we use the general expression for the tunneling transfer operator $\hat{T}$ in terms of the time-ordered exponent: \cite{T}
\begin{equation}
\hat{T}=\frac{d}{dt}\hat{T}_t e^{-i\int_{-\infty}^t dt' H_t(t')} |_{t=0},
\label{timerep}
\end{equation}
which gives the familiar expression, when expanded to second order in tunneling Hamiltonian in the energy domain,
\begin{equation}
I=2\pi \sum_{m}|\langle m|\hat{A}_L^\dagger \hat{R}_{0}\hat{A}_R|0\rangle|^2\delta(E_m-E_{0}),
\end{equation}
where $\hat{R}_{0}\equiv (E_{0}-H_0-H_d+i0)^{-1}$ is the retarded resolvent. However, in the present context it is convenient to keep the time representation (\ref{timerep}) and replace the delta-function by an integral over time. As a result, we arrive at the following expression:
\begin{equation}
\label{g}
I=\int_{-\infty}^\infty dt \int_{-\infty}^t dt_1 \int_{-\infty}^0 dt_2 \langle \hat{A}_L^\dagger(t_1) \hat{A}_R(t) \hat{A}_R^\dagger(0) \hat{A}_L(t_2)\rangle,
\end{equation}
where the averaging is taken over the ground state.

Next, we substitute tunneling operators (\ref{A}) into the Eq.\ (\ref{g}) and use the gaussianity of the fields $\phi_a$ to evaluate the average. Since the fluctuations of fields in different channels are independent, the average splits in the product of the four-point correlators of the form
\small
\begin{eqnarray}
\langle e^{i\xi\phi_a(t_1)}e^{i\zeta\phi_a(t)}e^{-i\zeta\phi_a(0)}e^{-i\xi\phi_a(t_2)}\rangle \nonumber \\ \propto \frac{e^{i\delta_{aL}\Delta\mu(\xi(t_2-t_1)-\zeta t)}}{(i(t_1-t_2)+0)^{\xi^2}(it+0)^{\zeta^2}}\nonumber \\ \times\frac{(i(t_1-t)+0)^{\xi\zeta}(-it_2+0)^{\xi\zeta}}{(it_1+0)^{\xi\zeta}(i(t-t_2)+0)^{\xi\zeta}}.
\end{eqnarray}
\normalsize
Here $\xi$ and $\zeta$ are arbitrary numbers and $\delta_{aL}$ is a Kronecker delta. The exponential oscillation function originates from the correlator of the fields $\phi_L$ and accounts for the fact that the left channel is biased.

Since the time evolution of the dot operators is trivial $\hat{d}(t)=e^{-i\epsilon t}\hat{d}$, their four-point correlator is just an oscillatory factor. Knowing the expressions for all the averages that appear in Eq.\ \eqref{g} and manipulating with the time variables and the integration contours, we arrive at the following expression for the cotunneling current
\begin{eqnarray}
\label{Gamma_in}
I \propto \int\limits_{-\infty}^\infty\!\! dt \! \int\limits_0^\infty\!\! dt' \!\! \int\limits_0^\infty \!\! dt''\frac{e^{i \Delta\mu t}e^{-\varepsilon(2t'+t'')}}{(2t' + t'' + it)^{1 - \beta}(it + 0)^{1 - \gamma}}\nonumber\\ \times\frac{t'^\alpha(t'+t'')^\alpha}{(t' + t''+ it)^{\alpha}(t' + it)^{\alpha}},
\end{eqnarray}
where the exponents are given by
\begin{equation}
\label{powers}
\alpha=(\beta+\gamma)/2,\;\beta=2\eta_L-\sum_a \eta_a^2,\;\gamma=2\eta_R-\sum_a \eta_a^2,
\end{equation}
and the parameter $\varepsilon$ is defined by the relation
\begin{equation}
\varepsilon\equiv \epsilon+\eta_L \Delta\mu -\Delta\mu.
\label{vare}
\end{equation}
This parameter represents the difference between the energy level of the dot, shifted by the voltage bias of the left channel, and the Fermi energy of this channel. The value of energy shift, $\eta_L \Delta\mu$, has the transparent physical meaning: The voltage bias, applied to the left channel, induces the extra electron density $\Delta\sigma_{a}(x)=\Delta\mu\int dx' V_{aL}^{-1}(x,x')$ in the channel $a$, which follows from the Eq.\ (\ref{screening}) with zeroes on the right hand side replaced by $\delta_{aL}\Delta\mu$. Replacing in the Eq.\ (\ref{remorm}) $\sigma_{a}$ with $\sigma_{a}+\Delta\sigma_{a}$ and using again the Eq.\ (\ref{screening}) with the definition of $\eta_L$, one arrives at the desired value of the energy shift. Thus $\varepsilon$ is an effective size of the energy gap for the virtual state in the cotunneling process, where the dot is occupied by an extra electron. 

Concerning the exponents (\ref{powers}) the following remark is in order. Depending on the experimental situation, it may happen that the fraction of electron charge $\eta_{g}\leq 1$ is screened by metallic gates surrounding the QD. Taking into account the fact that the Fermi wave length in metals is much smaller than the one in semiconductors, it is natural to assume, that the number $n_g$ of free-electronic channels in the metallic gates that contribute to screening of the dot is large. Assuming further, that the individual contributions $\eta_i$ from these channels are of the same order and using the condition $\sum_i \eta_i=\eta_g$, one arrives at the conclusion that the contribution of the gates to the Anderson part of the FES exponent, $\sum_i \eta_i^2$, scales to zero as $1/n_g$ and can be neglected. Thus, in reality all the induced charges in the semiconductor channels, $\eta_a$, do not necessary add to 1.

\begin{figure}[h]
	\centering
		\includegraphics[scale=0.3]{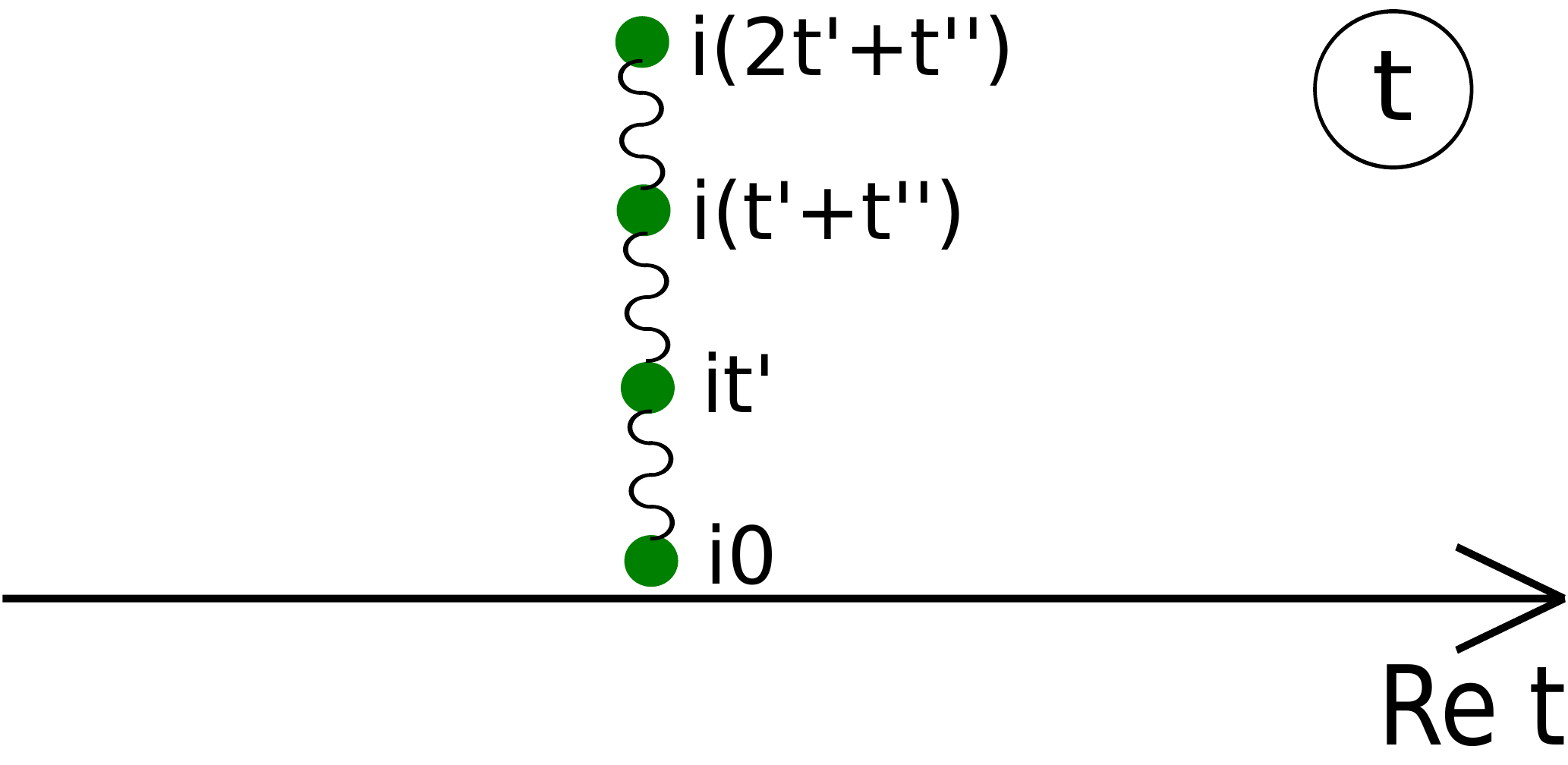} 
	\caption{The analytical structure of the expression under the integral (\ref{Gamma_in}) as a function of the variable $t$ is schematically shown.  The curly line represents the branch cut that connects four branching points.}
\label{fig:complexplain}
\end{figure}

All the integrals in Eq.\ (\ref{Gamma_in}) are convergent, since each time variable enters the integral with negative power and is integrated either with the oscillatory factor or with the exponential function, which decay rapidly at infinity.  The analytical structure of the integrand as a function of $t$ is presented in Fig.\ \ref{fig:complexplain}. It has one branch cut that connects four branching points, which follows from the fact that the total power of the denominator as a function of $t$ is integer. The integration over $t$ from $-\infty$ to $\infty$ allows one to deform  the contour, and thus simplifies calculations.

In the case of free fermions, i.e., where $\eta_a=0$ for all $a$, the integral (\ref{Gamma_in}) can be calculated exactly, leading to the well-known expression for the cotunneling current
\begin{equation}
\label{Gfree}
I \propto \frac{\Delta\mu}{\varepsilon(\varepsilon+\Delta\mu)},
\end{equation}
where, obviously, $\varepsilon=\epsilon-\Delta\mu=\epsilon_0-\Delta\mu$.
Below, we will use this result to compare to the interacting case. In general, the evaluation of the integral (\ref{Gamma_in}) as a function of two parameters, $\Delta\mu$ and $\varepsilon$, presents a challenge, and does not seem to be instructive. Indeed, only in the limits of the small bias, $\Delta\mu\ll\varepsilon$, and large bias, $\Delta\mu\gg\varepsilon$, one approaches the thresholds in the spectrum associated with two biased Fermi seas and thus expects the universal power-law behavior of the current. Therefore, in the rest of the paper we will concentrate on finding the asymptotic forms of the integral (\ref{Gamma_in}).

\section{FES exponents}

We start with the low-bias regime and analyze the asymptotic behavior of the integral  (\ref{Gamma_in}) for $\Delta\mu\ll\varepsilon$. Note, that in this regime $\varepsilon\approx\epsilon$, and we neglect the difference. We expect the linear dependence of the current on $\Delta\mu$, because in this regime the interactions are effectively screened. Indeed, in the course of electron tunneling from one channel to another, that occurs on the time scale $1/\Delta\mu$, quantum fluctuations of charge on the QD with characteristic time scale $1/\varepsilon$ are relatively fast and average to negligible values. Thus, cotunneling in this case can be considered as tunneling of an electron through an effective potential barrier,
that only depends on the energy gap $\varepsilon$.  

In order do find the asymptotic form of the integral (\ref{Gamma_in}), we integrate separately ``fast'' and ``slow'' functions. We note that the exponential function limits the integrals over variables $t'$ and $t''$ to small intervals of the size of the order of $1/\varepsilon$ around zero. On the other hand, slowly oscillating function of $t$ limits the integral over $t$ to large interval of the size of the order of $1/\Delta\mu$. Therefore, taking into account the relation (\ref{powers}) that connects the exponents, and keeping leading order terms in $\Delta\mu/\varepsilon$, we replace four purely imaginary branch points in the complex plane of $t$ (shown in Fig.\ \ref{fig:complexplain}) with one pole of the power 2. Evaluating the contribution of this pole, we obtain the integral:
\begin{equation}
\label{Gsmall1}
I\propto \Delta\mu\int\limits_0^\infty dt'\int\limits_0^\infty dt'' e^{-\varepsilon(2t'+t'')}t'^\alpha(t'+t'')^\alpha,
\end{equation}
which readily gives us the final expression
\begin{equation}
\label{Gsmall}
I\propto\frac{\Delta\mu}{\varepsilon^{2+\beta+\gamma}},\quad \Delta\mu\ll\varepsilon .
\end{equation}
Here, we again used the relation (\ref{powers}) in order to express the power-law function of $\varepsilon$ in terms of the exponents $\beta$ and $\gamma$, associated with two channels that are tunnel coupled to the QD.   We note that Eq.\ (\ref{Gsmall}) for $\beta=\gamma=0$ coincides with the free-fermionic expression (\ref{Gfree}) for small biases.

Next, we concentrate on the regime of the large bias, where the energy gap becomes small,  $\varepsilon\ll \Delta\mu$, and one approaches another threshold in the spectrum associated with the Fermi level of one of the leads. Taking the advantage of fast oscillations in the integral (\ref{Gamma_in}), we deform the contour of integration over the variable $t$ and observe, that the integral is limited to the interval $t\sim 1/\Delta\mu$ away from the origin. Since the decay of the integrand as a function $t'$ and $t''$ is determined by the parameter $\varepsilon$, and thus it is much slower, we conclude, that between four brunch points shown in Fig.\ \ref{fig:complexplain}, only one at $t=0$ contributes to the integral. This allows us to evaluate the integral over $t$ by neglecting the exponential function $e^{-\varepsilon(2t'+t'')}$ and then restoring it for the integral over variables $t'$ and $t''$. The integrals become trivial, and we obtain the new power-law behavior:
\begin{equation}
\label{Glarge}
I\propto\frac{1}{\Delta\mu^\gamma\varepsilon^{1+\beta}},\quad \Delta\mu\gg\varepsilon .
\end{equation}    
The non-interacting limit of this result agrees with the expression (\ref{Gfree}) for free fermions at large biases.

Interestingly, $\beta$ and $\gamma$ are exactly the FES exponents for the sequential tunneling to the QD from the corresponding channels, and vice versa. For instance, the rate of the sequential tunneling from the left and right lead scales as $1/\varepsilon^\beta$ and $1/\varepsilon^\gamma$, respectively, where this time $\varepsilon$ denotes the excess energy of an electron, i.e., we replaced $\varepsilon\to-\varepsilon$. Thus, in the cotunneling regime this implies the multiplicative FES effect from both leads, and the extra factor of $1/\varepsilon$ for each virtual transition. Finally, we note, that by doing the measurements of the current as a function of $\Delta\mu$ and $\varepsilon$ in the regime of large bias, one can extract both FES exponents, $\beta$ and $\gamma$. It would be then interesting to compare these results with the measurements of the current as a function of $\varepsilon$ in the low bias regime, in order to verify our theory.

\section{Conclusion}

The FES effect manifests itself as a power-law singularity in the density of states of tunneling to a QD at Fermi level of the metallic leads, and presents a textbook example of the phenomenon originating from interactions that cannot be accounted perturbatively. The key feature of the effect is the fact that the exponents of the power-law singularity are universal, since they depend only on the screening charges accumulated in the leads when one adds an electron or hole to the QD.

 Despite being thoroughly studied in various systems both experimentally and theoretically, this effect has recently received an attention in the context of the systems, where the number of Fermi edges is more than one.\cite{Mirlin} Namely, it has been proposed to attach to a QD system in the FES regime an addition voltage source that irradiates the QD with the shot noise.\cite{Iurii} This leads to a number of new interesting effects of essentially non-equilibrium character. In the present work,  we have proposed an alternative approach, based on cotunneling through a QD. This process does not perturb the leads, so they retain they equilibrium state. 

Since two electrons participate in cotunneling through a QD in the Coulomb blockade regime,  two Fermi edges are involved, and consequently two energy scales: potential difference and the Coulomb energy gap. We investigate this problem using the bosonization approach, which accounts the Coulomb long-range interactions exactly. We consider two limits of low and large bias, and find new power-law exponents in these regimes.  It turns out that they present a linear combination of ``classical'' FES exponents for the sequential tunneling, so that the contribution of two leads to the FES effect has a multiplicative character. We suggest an experiment where our predictions can be verified without any fitting parameter.

%

\begin{acknowledgments}
This work has been supported by the Swiss National Science Foundation.
\end{acknowledgments}

\end{document}